\documentclass[np2,twoside,onecolumn,showpacs]{revtex4}

\usepackage{amssymb}
\usepackage{amsmath}
\usepackage{graphicx, color, ulem}
\usepackage{bm,multirow}

 \volumenumber{xx} \issuenumber{x}
\volumeyear{Month Year} 
\startpage{1}
\endpage{3}
 
\def\S{{\bf S}}

\def\P{{\bf P}}

\def\centeron#1#2{{\setbox0=\hbox{#1}\setbox1=\hbox{#2}\ifdim
   \wd1>\wd0\kern.48\wd1\kern-.48\wd0\fi
   \copy0\kern-.48\wd0\kern-.48\wd1\copy1\ifdim\wd0>\wd1
   \kern.48\wd0\kern-.48\wd1\fi}}

\newcommand{\beq}{\begin{equation}}
\newcommand{\eeq}{\end{equation}}
\newcommand{\bea}{\begin{eqnarray}}
\newcommand{\eea}{\end{eqnarray}}
\newcommand{\ba}{\begin{array}}
\newcommand{\ea}{\end{array}}

\newcommand{\p}{\partial}
\newcommand{\nn}{\nonumber}

\begin{document}

\title{Gravitational Deflection of Light: A Heuristic Derivation at the Undergraduate Level}

\author{Hongbin \surname{Kim}}
\affiliation{Department of Physics Education, Pusan National University, Busan 46241, Republic of Korea\\
Research Center for Dielectric and Advanced Matter Physics, Pusan National University, Busan 46241, Republic of Korea \\
Department of Physics Education, Seoul National University, Seoul 08826, Republic of Korea}

\author{Dong-han \surname{Yeom}}
\email{innocent.yeom@pusan.ac.kr} 
\affiliation{Department of Physics Education, Pusan National University, Busan 46241, Republic of Korea\\
Research Center for Dielectric and Advanced Matter Physics, Pusan National University, Busan 46241, Republic of Korea}

\author{Jong Hyun \surname{Kim}}
\affiliation{Department of Physics Education, Pusan National University, Busan 46241, Republic of Korea\\
Research Center for Dielectric and Advanced Matter Physics, Pusan National University, Busan 46241, Republic of Korea}


\date[]{Submitted 23 December 2023}

\begin{abstract}
In this paper, we present a new heuristic derivation of the gravitational deflection of light around the Sun at the undergraduate level. Instead of solving the geodesic equation directly, we compute the correct deflection angle by focusing on the acceleration term of null geodesics. Using this heuristic deviation, we expect that undergraduate students who have not learned general relativity will be able to experience this computation, which is one of the most remarkable evidences of general relativity.
\end{abstract}


\maketitle

\section{Introduction}
 The gravitational deflection of light around the Sun is one of the striking evidences of the general theory of relativity (GR). In 1919, on the island of Príncipe, an observational expedition team led by Eddington confirmed that light passing around the Sun was deflected by 1.75 arcsec, as predicted by Einstein. Since the deflection of light was an important and dramatic observation that supports the validity of general relativity, it became one of the examples that must appear when introducing general relativity not only at the college level but even at the high school level. 
 
In order to teach this phenomenon to students who have not learned GR properly, qualitative explanations such as `the metaphor of a trampoline' \cite{Kersting, Kaur2017} or intuitive models for curved space are used \cite{Zahn2014}. If students want to obtain a quantitative result of 1.75 arcsec, they should be able to handle complex concepts such as the geodesic equation and the Christoffel symbols that usually appear in GR textbooks \cite{MTW1973, dInverno1992, Weinberg1972, Schutz1985, Hartle2003, Carroll2004}. However, these are not easy concepts for undergraduate students in usual college curriculum. Particularly, undergraduate students even majoring in physics often experience a barrier between qualitative explanations and quantitative calculations in GR, even though they already know quite a bit of the basic concepts of GR.

In order to overcome this difficulty, there have been some suggestions that provide quantitative results on the gravitational deflection angle of light through simplified calculations \cite{Mahajan2021, Pinochet2020, Lerner1997, Kassner2015}. For example, a study used the dimensional analysis and obtained a half of the observed value using the assumption that light is accelerated only in the near zone that corresponds to the diameter of the Sun \cite{Mahajan2021}. Another study presented the deflection angle through the comparison with the situation in which the spacecraft is accelerated \cite{Pinochet2020}. In this approach, the width of the spacecraft was chosen to be the diameter of the Sun to yield a half of the value, too, which is similar to the approach of Ref.~\cite{Mahajan2021} in that the parameter was chosen arbitrarily. Meanwhile, there were also approaches that yield the deflection angle by using the fact that the refractive index varies with distance from the Sun \cite{Lerner1997, Kassner2015}. In fact, this is essentially the same approach as Einstein showed in his 1916 paper \cite{Einstein1916a}. Although these approaches could help undergraduate students to focus on the physical meaning rather than just mathematical calculation, they still include some computational complexities such as expressing metrics in isotropic form. These are bound to be burdensome for average undergraduate students. 

In this paper, we propose a new heuristic derivation for the deflection angle of light passing around the Sun. In Section 2, the attempts to explain the gravitational deflection of light in Newtonian framework and Einstein's early approach are examined. This section will provide how we approximate the calculations.  In Section 3, we present a heuristic derivation at the undergraduate level. Finally, we summarize and discuss our results in Section 4.

\section{Historical context: from Newton to early Einstein}
Contrary to popular belief that ``light has no mass and therefore is not affected by gravity", Isaac Newton himself seriously considered the question of whether light would be affected by an object, and left this issue to be solved in the future. In 1704, Newton raised the following question at the {\it{Opticks}} (Query 1 appeared at the end of the book) \cite{Newton}:

``Do not Bodies act upon Light at a distance, and by their action bend its Rays, and is not this action strongest at the least distance? (Query 1)"

According to Newtonian mechanics, every object with non-zero mass experiences an inverse square force from the Sun. Moreover, the trajectories of the objects are independent of their mass provided that their initial velocities are the same. The reason is that the values of inertial mass and gravitational mass are equivalent. Mathematically, instead of assuming the mass of light to be zero, if one regards that the mass of light {\it{converges to zero}}, it might not be surprising even the light ray bends around the Sun in the same way as other ponderable objects. It is not surprising that Newton, who thought that light was made up of tiny discrete particles (called {\it{corpuscles}}), had this idea. But Newton himself would have had to leave the above question as an open question because he had no way of confirming whether the mass of light was absolutely zero or not.

The trajectory of an object is determined according to the inverse square law of gravitation and the Newton's second law of motion as follows:
\beq
a_r=\frac{GM}{r^2}.
\eeq
The above equation shows that the trajectory is independent of the mass of the moving object. Figure 1 shows how the object is deflected by the Sun through gravitation.
\begin{figure}[htp]
\begin{center}
	\includegraphics[scale=0.5]{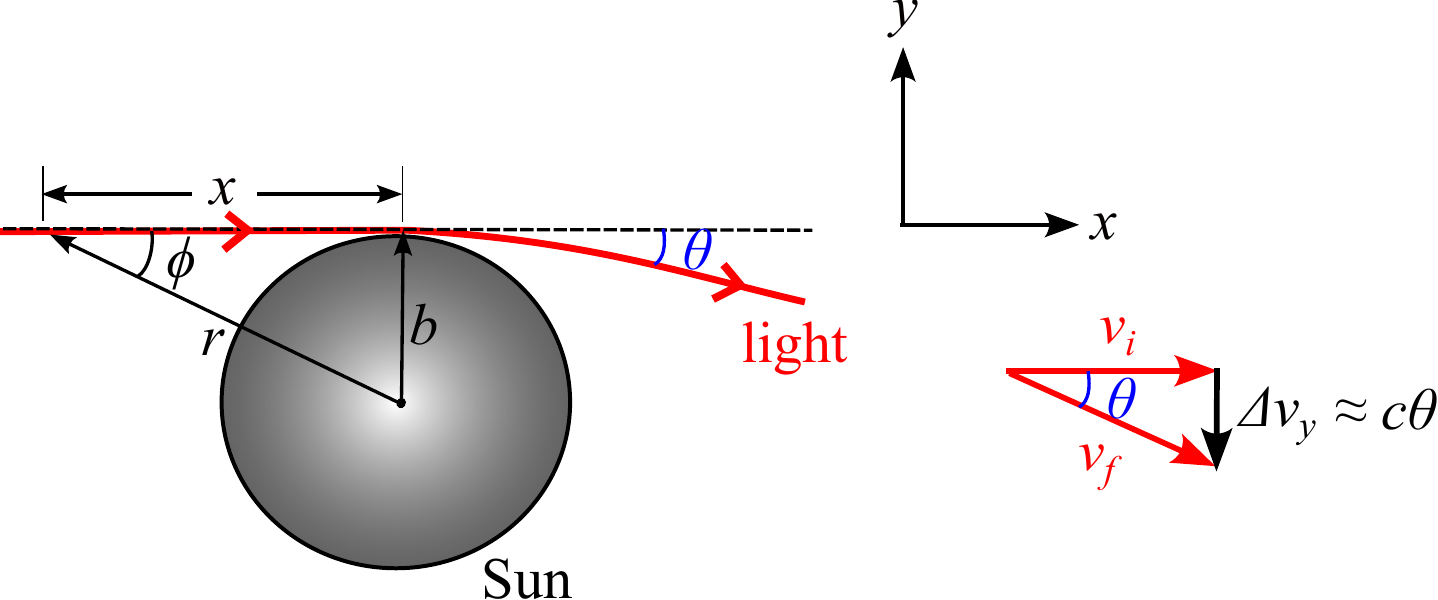}
	\caption{A conceptual figure for the light deflection near the Sun. A light ray goes from left to right, where the bending angle is defined by $\theta$. The initial speed is $v_{i}$ and the final speed is $v_{f}$; the magnitude of the difference of velocities $\Delta v_{y}$ is approximately $c\theta$, because $v_{i} = v_{f} = c$.} \label{figLlabel}
\end{center}
\end{figure}

Because the initial speed $v_i$ and the final speed $v_f$ of an moving object is equal to the speed of light $c$, and the deflection angle $\theta$ is very small (we assume that the impact parameter $b$ is much larger than the Schwarzschild radius $2GM/c^{2}$), the magnitude of the change in velocity of an object can be approximated as
\beq
\Delta v_y \approx c\theta.
\eeq
Meanwhile, the $y$-component of the change in velocity can be obtained by integrating the $y$-component of the acceleration as follows:
\bea
\vert \Delta v_y \vert &=& \int_{-\infty}^{\infty} \vert a_r \vert \sin\phi dt  \nn \\
&=&  \int_{-\infty}^{\infty} \frac{GM}{b^2+x^2} \times \frac{b}{\sqrt{b^2+x^2}} \times \frac{dt}{dx}dx \nn \\
&=& \frac{2GM}{cb},
\eea
where $b$ is the impact parameter, and $dx/dt \approx c$ is used because $\theta$ is very small. Hence, we obtain the deflection angle from $\theta=\vert \Delta v_y \vert/c$ as follows:
\beq
\theta=\frac{2GM}{c^2 b}=\frac{R_s}{R} \approx 0.87^{\prime\prime},
\eeq
where $R_s=2GM/c^2$ is the Schwarzschild radius of the Sun, and $b \approx R$, the radius of the Sun. This is half of the value that we have observed.

Already in 1804, Johann Georg von Soldner calculated the deflecting angle of light passing around the Earth using the method introduced above (seemingly different, but essentially the same), and obtained one-half of the general relativistic value~\cite{Soldner}. About 20 years earlier, around 1784, Henry Cavendish also obtained the similar result (not exactly the same as Soldner's result, but they agree at the first order of approximation), although he never published the result~\cite{Will1988, Cavendish}.

In 1911, almost one hundred years after Soldner's calculation, Albert Einstein obtained a similar result even before establishing the complete framework of GR~\cite{Einstein1911}. Einstein considered the equivalence principle and his famous formula about the photon energy, $E=h\nu$, and supposed the effective mass of a photon $m=E/c^2=h\nu/c^2$ by using the mass-energy equivalence, $E=mc^2$. But this argument gives still one-half of the relativistic value, 0.87 arcsec. Although Einstein's approach was a step forward from the Newtonian approach, he could not reach the exact result because he did not fully understand the role of the spacetime metric (i.e., gravitational field)\cite{Lerner1997}.

From a pedagogical  point of view, it will be helpful for students to introduce the above historical context from Newton's original query to Einstein's approach by only using Newtonian mechanics and the special theory of relativity before introducing general relativistic corrections.

\section{Relativistic correction: A heuristic approach}
One of the biggest difficulties faced by students who have not yet learned general relativity is to understand the fact that general relativity is a field theory. In other words, although Newtonian mechanics also deals with continuous objects, it is basically a theory that describes the motion of point particles. That is, continuous objects can be reduced to the collection of point particles in Newtonian mechanics. Meanwhile, GR is a field theory that describes the gravitational phenomena using the ‘fields’ concept\footnote{The concept of `fields' can be used in various meanings depending on the context. In this paper, we used the term `fields' as a dynamical variable as a function of space and time. In that sense, so called the gravitational fields in Newtonian mechanics is not referred to as a field in this article since they describe only static situations.} instead of ‘forces’. Although the particle concept also appears in field theory and sometimes the interaction between particles and fields are of interest, basically the dynamical variables of a field theory is the fields. And fields are not reduced to particles in field theory. In GR, the gravitational field is expressed as a spacetime metric. 

In typical undergraduate curriculum in physics, students learn Maxwell's electromagnetic theory, the prototype of classical field theory, so they can get acquainted with the field concept. However, unlike the electric fields and magnetic fields (those can be usually expressed as vector fields), the gravitational field is usually described by a tensor fields of rank two. Of course, second rank tensor is not a completely unfamiliar concept to undergraduate students, because they also appears in electromagnetic theory (even in Newtonian mechanics, for example, tidal tensor or strain tensor etc.). But they are not the main dynamical variables both in electromagnetic theory and Newtonian mechanics. Therefore, it seems reasonable to assume that a second rank tensor (as a field) is a relatively unfamiliar concept to average undergraduate students.

One way to overcome this difference between general relativity and Newtonian mechanics is to emphasize the role of spacetime metrics in the classical mechanics course. In other words, by simply introducing the spacetime metric in a classical mechanics course and explaining its role through a simple example instead of introducing the second rank tensor in details, the barrier that students feel in learning GR can be significantly lowered. So called the \textit{Geometrized Newtonian Formulation} is a good example of this strategy \cite{Hartle2003}. Students who have learned the Lagrangian formulation and the variational method are expected to know that one can derive the Newtonian equation of motion from the following action:
\beq
S=-mc\int ds,
\eeq
where $m$ is the mass of a particle\footnote{In this article, we consider the motion of light, i.e., a massless particle. Hence, in this case, $m$ is to be understood as just a formal parameter of mass dimension. The value of $m$ does not affect the final result.}, $c$ is the speed of light, and $ds$ is the line element ({\it{distance}} concept in spacetime) that determines the world-line of a particle in spacetime \cite{Landau1976}. For a free particle, the line element is given as a Minkowski metric in the following form:
\beq
ds^2=-c^2 dt^2+dx^2+dy^2+dz^2.
\eeq

If the potential $\Phi$ is turned on, the equation of motion of a particle, $-m\vec{\nabla}\Phi=m\vec{a}$, can be expressed in geometric terms by using the line element (see Ref.\cite{Hartle2003})
\beq
ds^2=- \left( 1+\frac{2\Phi}{c^2} \right)c^2 dt^2+ \left(1-\frac{2\Phi}{c^2} \right) \left( dx^2+dy^2+dz^2 \right).
\eeq
This is the first step of our approach. It is important to have students to know that the concept of spacetime metric is not an exclusive property of the general relativity. In other words, spacetime metric can be introduced in the undergraduate-level classical mechanics course as an another formalism of mechanics. In this framework, one can derive the equation of motion by applying the variational method with the action Eq. (5) and the line element Eq. (7) and substituting $\Phi$ by $-GM/r$ for a point-like mass distribution.

The second step is to introduce the Schwarzschild metric as a solution to the field equation that describes the spacetime around the Sun. By substituting $-GM/r$ in place of $\Phi$ in Eq.~(7) and adding some explanations, we obtain the Schwarzschild metric as follows:\footnote{Of course, the coefficient of the $dr^2$ term in Eq.~(8) cannot be derived directly from Eq. (7) by this substitution. The inverse is possible when $r \gg GM/c^{2}$ is assumed}. From a teaching and learning perspective, there might be two strategies: either one uses a hand-waving argument to justify the Schwarzschild metric, or one exploits Eq.~(7) and obtain Eq.~(14) using perturbative expansions. The answer to which strategy is better depends on the teaching and learning situation.
\beq
ds^2=- \left( 1-\frac{2GM}{c^2 r} \right) c^2 dt^2+\frac{1}{1-\frac{2GM}{c^2 r}}dr^2+r^2 d \bar{\theta}^2+ r^2\sin^2 \bar{\theta} d \varphi^2.
\eeq
Once the above Schwarzschild metric is obtained, it may be possible to visualize how the spacetime is curved by using embedding diagrams with $\bar{\theta}=\pi/2$ slice \cite{MTW1973}.

The third step is to introduce the trajectories of light in the geometrized Newtonian formulation. In general, defining a light ray is not an easy task, but in the context of general relativity, light can simply be defined as a physical object that satisfies the null condition, $ds^2=0$. We may help students to remind that the condition, $ds^2=0$, corresponds the world-line of the light in special relativity. Then the trajectory of light is given by
\beq
0=-\left( 1-\frac{2GM}{c^2 r} \right)c^2 \dot{t}^2+\frac{1}{1-\frac{2GM}{c^2 r}}\dot{r}^2+ r^2\dot{\varphi}^2,
\eeq
where $\dot{}$ represents the derivative with respect to the affine parameter. 

The fourth step is to applying the variational method. In order to use the Lagrangian formalism, we set the Lagrangian describing the path of light as follows:\footnote{This form of Lagrangian amounts to $ds^2$ if you exclude the affine parameter. At this stage students may ask why we choose $ds^2$ as a Lagrangian instead of $ds$ as appeared in Eq. (5). It should be noticed here that if the Lagrangian is not an explicit function of the affine parameter, $L$ and $L^2$ yield the same equations of motion.}
\beq
L(t, \dot{t}, r, \dot{r}, \varphi, \dot{\varphi})=- \left(1-\frac{2GM}{c^2 r}\right)c^2 \dot{t}^2+\frac{1}{1-\frac{2GM}{c^2 r}}\dot{r}^2+ r^2\dot{\varphi}^2,
\eeq
and after obtaining the equation of motion, we will substitute $L=0$ again. Students need to be reminded that the above Lagrangian is a function of six variables: $t$, $\dot{t}$, $r$, $\dot{r}$, $\varphi$, and $\dot{\varphi}$. But as is well known in classical mechanics, the variables $t$ and $\varphi$ are cyclic. So, their conjugate momenta are conserved as follows:
\bea
p_t &=& \frac{\p L}{\p \dot{t}}=-2 \dot{t} \left(1-\frac{2GM}{c^2 r}\right)c^2  \equiv -2c^2 E, \\
p_\varphi &=& \frac{\p L}{\p \dot{\varphi}}=2r^2 \dot{\varphi} \equiv \frac{2l}{m},
\eea
where, we formally introduced $m$, which has the mass dimension, and $l$, which has the angular momentum dimension, to treat this problem as if describing the motion of a massive particle. And the constant value $E$ is defined as $-p_t/2c^2$. The derivative in the above equations is actually the derivative with respect to the affine parameter. But students can consider this derivative with respect to ``time" in the analogy of classical mechanics, because the problem-solving procedure resembles typlical central-force problem which is familiar to undergraduate students. By substituting $E$ and $l$ into Eq.~(10) and setting $L=0$ again, we obtain the following equation:\footnote{Of course, to be precise, one must follow the order of substituting the conserved quantities $p_t$ and $p_\varphi$ after calculating the Hamiltonian or the Routhian, but in this case, there is no difference.}
\beq
\dot{r}^2+ \left( 1-\frac{2GM}{c^2r} \right) \frac{l^2}{m^2 r^2}=E^2.
\eeq
At this stage, one could plot the effective potential experienced by light. Students could recognize Eq. (13) as a kind of energy conservation relation by comparing this with the energy conservation in the central-force problem.

Finally, by differentiating Eq.~(13) with respect to the affine parameter, we obtain the equation of motion as follows:
\beq
\ddot{r}=\frac{l^2}{m^2 r^3}-\frac{3GMl^2}{m^2 c^2 r^4},
\eeq
where $l$(=$mcb$) is the angular momentum of light (See Fig. 1). Therefore, from the properties of the polar coordinates, the radial component of the acceleration is
\beq
a_r=\ddot{r}-r\dot{\varphi}^2=\ddot{r}-\frac{l^2}{m^2 r^3}=-\frac{3GMl^2}{m^2 c^2 r^4}.
\eeq
Unlike the functional dependence of the acceleration on $r$ in Eq.~(1), this is $1/r^4$ type, not $1/r^2$. By substituting Eq.~(15) into $a_r$ in Eq.~(3), we obtain $\vert \Delta v_y \vert$ as follows:
\bea
\vert \Delta v_y \vert &=& \int_{-\infty}^{\infty} \vert a_r \vert \sin\phi dt \nn \\
&=& \frac{3GMl^2}{m^2 c^2} \int_{-\infty}^{\infty} \frac{1}{(b^2+x^2)^2} \times \frac{b}{\sqrt{b^2+x^2}} \times \frac{dt}{dx}dx \nn \\ 
&=& \frac{4GM}{cb}.
\eea
Hence, we finally obtain the deflection angle
\beq
\theta=\frac{\vert \Delta v_y \vert}{c}=\frac{4GM}{c^2 R}=\frac{2R_s}{R} \approx 1.75^{\prime\prime}.
\eeq
This result is in exact agreement with the observation.

In this approach, we tried to utilize the Newtonian type of equation of motion as much as possible and minimize the steps in which the approximations are used. As shown in Eq.~(2), instead of considering the whole trajectory, we focused on obtaining the acceleration term (i.e., the change of the velocity) as the minimal information to obtain the deflection angle.  A comparison to other approaches is shown in Fig.~2.

\begin{figure}[htp]
\begin{center}
	\includegraphics[scale=0.81]{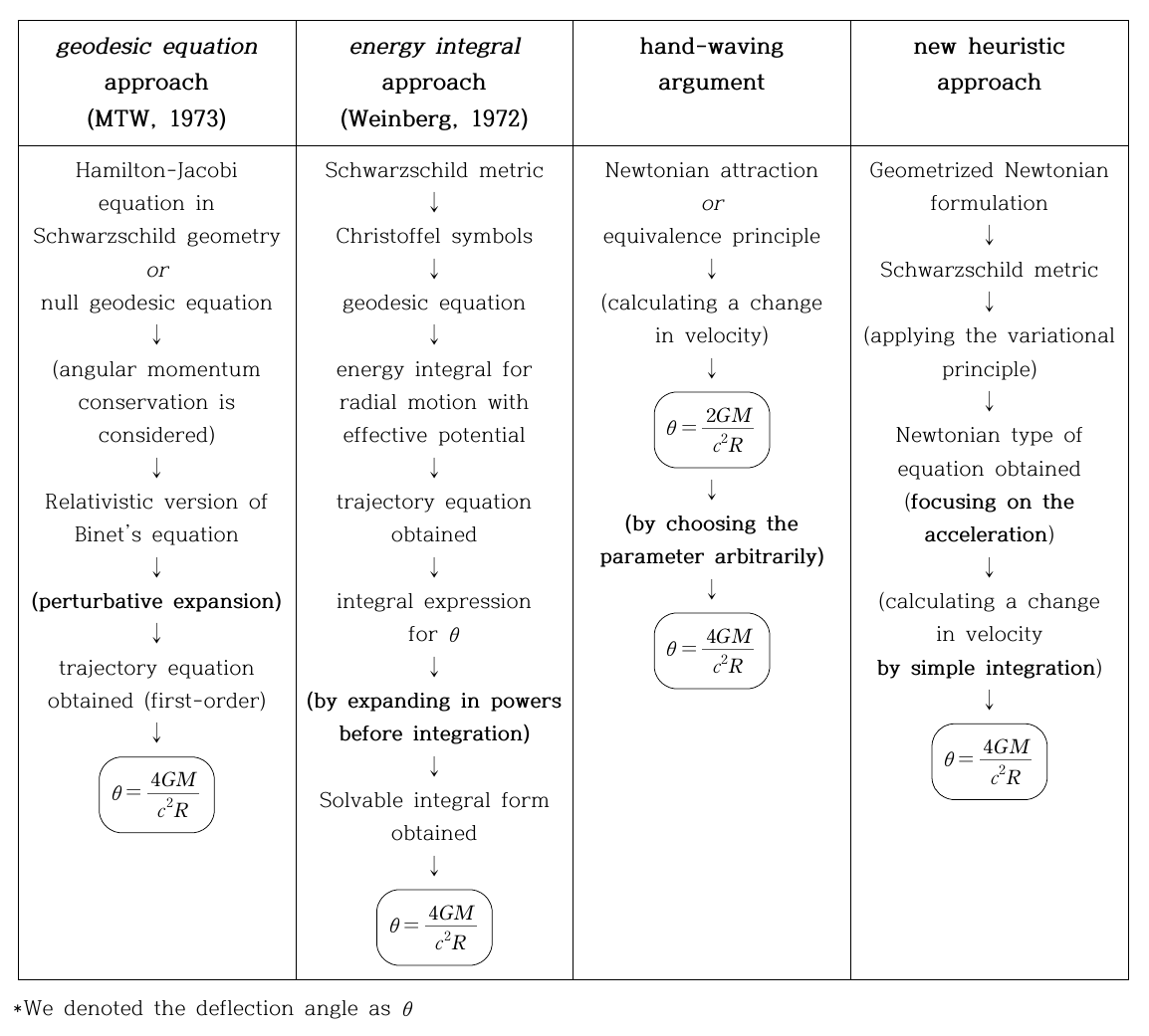}
	\caption{Comparison to other approaches: the geodesic equation approach, the energy integral approach, the hand-waving argument, and our new heuristic approach.} \label{tableLlabel}
\end{center}
\end{figure}

The approaches shown in typical GR textbooks are largely divided into two groups: the geodesic equation approach \cite{MTW1973, dInverno1992} and the energy integral approach \cite{Weinberg1972, Schutz1985, Hartle2003, Carroll2004}. In the geodesic equation approach (a representative example of this type is MTW's landmark book \cite{MTW1973}), a second-order differential equation (relativistic version of the Binet's formula) that describes the whole trajectories of light is derived, and then one needs to solve the perturbed equations order by order. On the other hand, in the energy integral approach (a representative example  of this type is Weinberg's book \cite{Weinberg1972}), the energy integral for radial motion with effective potential describing whole trajectories of light is obtained, and then the deflection angle is expressed in the integral form. Here, a delicate power series expansion is necessary to make the expression integrable.

Both of these approaches must be difficult calculations at the elementary level undergraduate course. For the pedagogical purposes, hand-waving arguments were suggested to avoid these complicated calculations. However, it is hardly to say that these arguments make the students experience real calculations because they use the arbitrary parameters {\it{by hand}} to double the final value \cite{Mahajan2021, Pinochet2020}. Our heuristic approach shows a third way that both avoids the hand-waving argument and complicated approximation techniques.

\section{Summary and Discussion}
In this study we presented a new heuristic derivation of the gravitational deflection angle of light around the Sun at the undergraduate level. From a pedagogical point of view, this approach is expected to provide more plausible calculation rather than simply doubling the result of Newtonian gravity as shown in several previous studies. Moreover, the computational details in this approach is much easier for undergraduate students to understand than directly solving the geodesic equation by using complicated approximations.

Of course, even in this approach, having to require the Schwarzschild metric as a minimal prerequisite could be a heavy burden on undergraduate students. However, as described in Section 3, if the role of the spacetime metric is introduced to some extent through classical mechanics courses, it is expected that the derivation presented in this study can be understood sufficiently without a detailed understanding of GR as a field theory. Even if it is explained to students who do not have prior knowledge on the Lagrangian formalism or the role of the spacetime metric in classical mechanics, the main calculation in this study can be introduced just by substituting the $1/r^4$ term instead of the inverse squared term, $1/r^2$, in the equation of motion as an alternative pedagogical strategy.

One of the advantages of the approach presented in this study is that approximations are already used from the first stage of the computation. Although the higher-order terms are ignored, e.g., Eq.~(2), they do not affect the final result on the deflection angle; this is the reason why our approach does not require complicated perturbative technologies. Moreover, another advantage is that it actively uses computational environments already familiar in classical mechanics, such as focusing on the acceleration term, Lagrangian, cyclic coordinates, effective potentials and familiar integrals of rational functions.

The authors expect that this study will be helpful to students who have college physics-level knowledge and instructors who have experienced difficulties in explaining the gravitational deflection quantitatively at the introductory level. Similar attempts can be made for easy understanding through quantitative calculations at the undergraduate level for other important observational evidences of GR, such as gravitational waves and perihelion precession of Mercury. We leave these topics for future research topics.

\section*{ACKNOWLEDGEMENTS}
The authors would like to thank the anonymous reviewers for their helpful comments. This work was supported by the 2-Year Research Grant of Pusan National University. 



\begin{thebibliography}{99}

\bibitem{Kersting}
M. Kersting and D. Blair D, {\it{Teaching Einsteinian Physics in Schools}} 1st ed (Routledge, New York, 2021), Chap.10.

\bibitem{Kaur2017}
T. Kaur, D. Blair, J. Moschilla, W. Stannard and M. Zadnik, Phys. Educ. {\bf 52}, 065012 (2017).

\bibitem{Zahn2014}
C. Zahn and U. Kraus,  Eur. J. Phys. {\bf 35}, 055020 (2014).

\bibitem{MTW1973}
C. W. Misner, K. S. Thorne and J. A. Wheeler, {\it{Gravitation}} (W. H. Freeman, San Francisco, 1973), MTW for abbreviation.

\bibitem{dInverno1992}
R. D'Inverno, {\it{Introducing Einstein's Relativity}} (Clarendon press, Oxford, 1992).

\bibitem{Weinberg1972}
S. Weinberg, {\it{Gravitation and Cosmology: Principles and Applications of the General Theory of Relativity}} (Wiley, New York, 1972).

\bibitem{Schutz1985}
B. F. Schutz, {\it{A First Course in General Relativity}} (Cambridge University Press, New York, 1985).

\bibitem{Hartle2003}
J. B. Hartle, {\it{Gravity: An Introduction to Einstein's General Relativity}} (Addison-Wesley, San Francisco, 2003).

\bibitem{Carroll2004}
S. Carroll, {\it{Spacetime and Geometry: An Introduction to General Relativity}} (Addison-Wesley, San Francisco, 2004).

\bibitem{Mahajan2021}
S. Mahajan, Am.~J.~Phys. {\bf 89} (8), 749 (2021).

\bibitem{Pinochet2020}
J. Pinochet, Phys. Edu. {\bf 55}, 065017 (2020).

\bibitem{Lerner1997}
L. Lerner, Am.~J.~Phys. {\bf 65} (12), 1194 (1997).

\bibitem{Kassner2015}
K. Kassner, Eur.~J.~Phys. {\bf 36}, 065031 (2015).

\bibitem{Einstein1916a}
A. Einstein, Ann. Phys., Lpz. {\bf 354}, 769 (1916).

\bibitem{Newton}
I. Newton, {\it{Opticks: or, a Treatise of the Reflexions, Refractions, Inflexions and Colours of Light}} 4th ed. (Dover, New York, 2012).

\bibitem{Soldner}
S. L. Jaki, Foundations of Physics {\bf 8}, 11 (1978).

\bibitem{Will1988}
C. M. Will, Am.~J.~Phys. {\bf 56}, 413-414 (1988).

\bibitem{Cavendish}
K-H. Lotze and S. Simionato, Ann. Phys. {\bf 534}, 2200102 (2022).

\bibitem{Einstein1911}
A. Einstein, Ann. Phys {\bf 340} (10), 898 (1911).


\bibitem{Landau1976}
L. D. Landau and E. M. Lifshitz, {\it{Mechanics}} 3rd Ed. (Butterworth-Heinemann, Oxford, 1976).

\end{thebibliography}
\end{document}